\documentstyle[preprint,aps]{revtex}
\begin{document}
\draft
\title{New Algebraic Equation determining Anomalies}

\author{Haewon Lee\cite{cbnu}}

\address{Department of Physics, University of Michigan \\
Ann Arbor, MI 48109, USA}

\maketitle

\begin{abstract}
We derive a new algeraic relation which can be used to
find various spinor loop anomalies.
We show that this relation includes the Wess-Zumino consistent condition.
For an example, we consider the chiral anomaly.
With this formalism, the consistent anomaly and the covariant anomaly
are determined simultaneously.
\end{abstract}

\narrowtext

In quantum field theories, the breakdown of local symmetries prevent
us from constructing a consistent theory.
In particular, fermion one-loop amplitudes have been the main source of such
anomalies. The functional determinants of the Weyl fermion is known to have two
kinds of anomalies, chiral anomaly\cite{bardeen}
and gravitational anomaly\cite{witten}.

Anomalies can be found by either direct calculations or
by solving some algebraic equations which anomalies should obey, i.e.,
the Wess-Zumino consistency condition\cite{wess}.
The descent equation technique\cite{zumino_wu,bonora} gives us an elegant way
to find the
solution of the Wess-Zumino consistency condition for both kinds of
anomalies in the general space-time dimension. However this method does not
determine
the overall constants, which can be determined by direct calculations.

When  Alvarez-Gaum\`e and Witten first discovered a gravitation anomaly
in their pioneering work,
they also presented a close form of chiral and gravitation
anomalies\cite{witten}.
Shortly after that work,
Bardeen and Zumino pointed out \cite{zumino} that the anomalies obtained by
Alvarez-Gaum\`e and Witten did not satisfy the Wess-Zumino consistency
condition,
and differed from the consistent anomalies by redefining the classical current.
The form of anomalies obtained by Alvarez-Gaum\`e and Witten are covariant
under
gauge and diffeomorphic transformations, and are now called ``covariant
anomaly''.

For the covariant anomalies,  analytic methods were usually used  ,
where functional determinants are the main target of study.
Recently, however,  there have been many approaches \cite{zhang,tsutsui,abud}
to understand the algebraic structure of
covariant anomalies.  In contrast , in finding the consistent anomalies,
the algebraic method using Wess-Zumino consistency condition was usually used.
This is because
the heat kernel expansions used to find anomalies have complicated forms in the
case of the consistent anomalies compared with the case of covariant anomalies.
Moreover, in the case of the consistent gravitational anomalies,
the calculation of the heat kernel expansion neccessary to calculate anomalies
is
a formidable one.
Therefore,  most of works on the consistent anomalies relied
on Wess-Zumino consistency condition.
In spite of these difficulties,
analytic calculations of consistent anomalies were
done by some authors in a restricted class of
theories\cite{langouche,leutwyler}.

In this paper we derive a new algebraic equation different from
the Wess-Zumino consistency condition. However this new relation can be shown
to
include the Wess-Zumino consistency condition.

The Weyl determinants in a curved $D$-dimensional
space-time is formally written as
\begin{equation}
i \Gamma =  \ln\mbox{Det}({\cal D}P) \label{det}, \label{weyl_det}
\end{equation}
where
${\cal D}  =  -i E^\mu_{\ m}\gamma^m
( \partial_{\mu} - i A_\mu - \frac{i}{4} \omega^{mn}_\mu \sigma_{mn} )$, and
 $\omega^{mn}_\mu$ is the spin connection.
In Eq.\ [\ref{weyl_det}], $A_{\mu}$ and $E_\mu^{\ m}$  denote the gauge fields
and  the vielbein fields, respectively.
We use $\mu$, $\nu$, ...
to denote Einstein indices, and $m$, $n$, ... to denote Lorentz indices.
The metric tensor
$g^{\mu\nu}$ equals to $E^\mu_{\ m} E^{\nu m}$.
We use the metric $\eta^{mn}=(1,\ldots,1,-1)$, and our Dirac matrices statisfy
$\{\gamma_m,\gamma_n\} = - 2 \eta_{mn}$. In Eq.[\ref{weyl_det}] ,
$\sigma_{mn} = \frac{i}{2}[\gamma_m,\gamma_n]$ , and
$P$ denotes one of $\{P_+, P_-\}$, where  $ P_\pm \equiv \frac{1 \pm
\gamma^{D+1}}{2}$.

The formal expression given in Eq.[\ref{weyl_det}] is ill-defined since
the operator maps one space to a different space.
So we use, for the definition of Weyl determinants,

\begin{equation}
i\Gamma  =  \ln\mbox{Det}({\cal D}_0{\cal D})_P ,
\end{equation}
where the subscript $P$ denotes the projection to the subspace corresponding to
$P$,
and ${\cal D}_0$ is some fixed Dirac operator.
For an example, we can choose ${\cal D}_0 = -i \delta^\mu_{\ m}\gamma^m
\partial_\mu$.

Next we seperate $\Gamma$ into two parts
\begin{equation}
 \Gamma = \Gamma_{\rm e} + \Gamma_{\rm o} \label{gamma},
\end{equation}
where
\begin{eqnarray}
i\Gamma_{\rm e} &=&  \frac{1}{2}
   \left( \ln\mbox{Det}({\cal D}_0{\cal D})_P +
 	\ln\mbox{Det}({\cal D}{\cal D}_0 )_P \right) , \nonumber \\
i\Gamma_{\rm o}  &=&  \frac{1}{2}
   \left( \ln\mbox{Det}({\cal D}_0{\cal D})_P -
 	\ln\mbox{Det}({\cal D}{\cal D}_0 )_P \right)  . \nonumber
\end{eqnarray}
Since we can write
   $i \Gamma_{\rm e} = \frac{1}{2} \ln\mbox{Det}({\cal D}{\cal
D})_{\overline{P}}$,
$\Gamma_{\rm e}$ is manifestly covariant for all kinds of symmetry
transformations,
and has no anomalous part. However $\Gamma_{\rm o}$ can have
anomalous part, and its variations respect to some symetry transformations give
so-called consistent anomalies.
To study $\Gamma_{\rm o}$ in a more general fashion,
let us consider a new functional
\begin{equation}
i \widetilde{\Gamma}_{\rm o}[{\cal D}',{\cal D}]  \equiv \frac{1}{2}
   \left( \ln\mbox{Det}({\cal D}'{\cal D})_P -
 	\ln\mbox{Det}({\cal D}{\cal D}' )_P \right) ,
\end{equation}
where ${\cal D}'$ corresponds to ${\cal D}$ with $E^{\nu m}= E'^{\nu m}$
and $A_\mu = A'_\mu$.
Note that $\widetilde{\Gamma}_{\rm o}[{\cal D}',{\cal D}]
= - \widetilde{\Gamma}_{\rm o}[{\cal D},{\cal D}'] $ and
$\Gamma_{\rm o}[{\cal D}] = \widetilde{\Gamma}_{\rm o}[{\cal D}_0,{\cal D}] $.
Consider the intrinsic parity operation \cite{witten1}
\begin{eqnarray}
 E^\mu_a  & \rightarrow & E^\mu_a \,\,\,\,\,\mbox{when}\,\,\,\, a \neq D , \\
\nonumber
E^\mu_D & \rightarrow & -E^\mu_D .
\end{eqnarray}
$E'^\mu_a$ transforms in the same way. Using the fact that
${\cal D}$ transforms to $- \gamma_D {\cal D} \gamma_D$,
one can easily show that
$\widetilde{\Gamma}_{\rm o}$ is odd under the intrinsic parity operation.

For the heat kernel representation of $\widetilde{\Gamma}_{\rm o}$, we use
\begin{equation}
i\widetilde{\Gamma}_{\rm o}[{\cal D}',{\cal D}] \equiv  - \frac{1}{2}
\int_{\xi}^{\infty}
		\frac{1}{\tau}\, d\tau\,
              \mbox{Tr}\, P(\,e^{ - \tau {\cal D}'{\cal D}} -
		\,e^{ - \tau {\cal D}{\cal D}'} ) . \label{func}
\end{equation}
where $\xi$ denotes the proper time cutoff and $\mbox{Tr}$ the functional
trace.
Next let us suppose that ${\cal D}$ and ${\cal D}'$ are parametrized by
$s$ and $s'$, respectively, so that
${\cal D}(0)={\cal D}_0$ , ${\cal D}(1)={\cal D}$,
${\cal D}'(0)={\cal D}_0$ , and ${\cal D}'(1)={\cal D}'$.
 Then one can prove that
\begin{equation}
i \frac{\partial^2}{\partial s\partial s'} \widetilde{\Gamma}_{\rm o}[{\cal
D}',{\cal D}]
  = - \frac{1}{2} \int_0^{\xi}
		\, d\tau\, \left[ \mbox{Tr}\, P\left(\,e^{- \tau {\cal D}'{\cal D}}
		\frac{d}{d s'}{\cal D}'\,e^{- (\tau-\xi) {\cal D}{\cal D}'}
		\frac{d}{d s}{\cal D} \, - \, ( {\cal D} \leftrightarrow {\cal D}' ) \right)
\right] . \label{two_var}
\end{equation}
Notice that the integration variable $\tau$ in Eq.[\ref{two_var}] varies from
$0$ to $\xi$.

Now we should take the limit $\xi \rightarrow +0$ in Eq.[\ref{two_var}].
We expect that small $\xi$-expansion similar to the heat kernel expansion
would exist in this case. This can be explictly proved in the case of flat
space-time
using the ordinary heat kernel expansions. Since small $\xi$ governs
the short distance behaviours of the operators,
we may assume that this holds in general cases.
So the right hand side of Eq.[\ref{two_var}], for an infinitesimal $\xi$,
is a parity odd local function of various fields appearing
in ${\cal D}$ and ${\cal D}'$. Integrating the both sides with respect to $s$
and $s'$, we
can write
\begin{equation}
\widetilde{\Gamma}_{\rm o}[{\cal D}',{\cal D}] = \Gamma_{\rm o}[{\cal D}]
	- \Gamma_{\rm o}[{\cal D}'] + h({\cal D}',{\cal D}) , \label{two_var1}
\end{equation}
where $h({\cal D}',{\cal D})$ is some function of odd parity which admits
the small $\xi$-expansions like $\xi^{- D/2}\sum_{i=0}^\infty \xi^i h_i$, where
$h_i$ are local functions.
Note that $h({\cal D}',{\cal D}) = - h({\cal D},{\cal D}')$,
and $h({\cal D},{\cal D}_0) = 0$.

Now we briefly review about the three kinds of local transformations,
i.e. , gauge transformations,
Lorentz transformations and generalized coordinate transformations.
The local gauge transformation is described by
\begin{equation}
\delta_\Lambda A_\mu = - \partial_\mu \Lambda + i [\Lambda,A_\mu] ,
\end{equation}
where $\Lambda$ denotes an infinitesimal generator of gauge transformations.
${\cal D} $ transforms as
$\delta_\Lambda {\cal D}  = [ i\Lambda, {\cal D} ]$ .
The local Lorentz transformation is described by
\begin{equation}
\delta_\theta e_\mu^{\ m}   =   \theta^m_{\ n}(x) e_\mu^{\ n} . \label{lorentz}
\end{equation}
One can easily show that, under the local Lorentz transformation,
  $\delta_\theta {\cal D} = [\frac{i}{4} \theta_{mn} \sigma^{mn},{\cal D}]$ .
An infinitesimal generalized coordinate transformation is given by
\begin{equation}
\delta_\eta e_\mu^{\ m} = \eta^\nu \partial_\nu e_\mu^{\ m}
	+ e_\nu^{\ m}\partial_\mu\eta^\nu  \label{einstein} .
\end{equation}
$A_\mu$ transforms in a similar way.
Under this transformation,
$\delta_\eta {\cal D} = [ \eta^\mu \partial_\mu , {\cal D} ]$.

Next let us consider the result of symetry transformation on
Eq.[\ref{two_var1}].
If we assume that ${\cal D}'$ transforms in the same way as ${\cal D}$ ,
$\widetilde{\Gamma}_{\rm o}[{\cal D}',{\cal D}]$ is manifestly covariant,
and we have
\begin{equation}
a_{cons}({\cal D}) - a_{cons}({\cal D}') + \delta h({\cal D}',{\cal D})
	=  0 , \label{anom_rela}
\end{equation}
where $ a_{cons}({\cal D}) \equiv \delta \Gamma_o[{\cal D}] $.
Eq.[\ref{anom_rela}] can be used to determine $a_{cons}$.
Using $h({\cal D},{\cal D}_0) = 0$, one can easily prove
\begin{equation}
a_{cons}({\cal D}) - a_{cons}({\cal D}')|_{_{{\cal D}' = {\cal D}_0}}
 = - \delta_{{\cal D}'} h({\cal D}',{\cal D})|_{_{{\cal D}' = {\cal D}_0}} ,
 \label{anom_relb}
\end{equation}
where $\delta_{{\cal D}'}$ denotes the infinitesimal transformation
restricted only to ${\cal D}'$.
Eqs.[\ref{anom_rela}-\ref{anom_relb}] are our main result.
Remember that Eq.[\ref{anom_rela}] holds for general $\xi$.
$a_{cons}$ also has a small $\xi$-expansion as
$\xi^{- D/2}\sum_{i=0}^\infty \xi^i (a_{cons})_i$ .
On the other hand,  we can show that, when $i \neq \frac{D}{2}$,
$(a_{cons})_i$ can be
written as $\delta F_i$, where $F_i$ is a some local function. $F_i$ is
universal in the sense
that it does not depend on the type of infinitesimal transformations $\delta$.
Here we omit the detailed analysis. Therefore if we subtract
$\xi^{- D/2}\sum_{i=0}^{\frac{D}{2}-1} \xi^i F_i$ from $\Gamma_o$
and take the limit $\xi \rightarrow +0$ , we have a subtracted generating
functional,
whose anomalies satisfy the same equation as Eq.[\ref{anom_rela}] with $h =
h_{D/2}$.
Notice that Eq.[\ref{anom_rela}] should be satisfied for all symmetry
transformations.
Eq.[\ref{anom_rela}] is linear in $h$ , and  satisfied again even if we
add to $h$ an arbitrary local function which is invariant
under all symmetry transformations.

If some local function $h$ with odd parity satisfies Eq.[\ref{anom_rela}] ,
then
the corresponding anomaly determined from Eq.[\ref{anom_relb}] obeys
the Wess-Zumino consistency condition automatically.
For an example, let us consider the case of gauge transformation.
Then Eq.[\ref{anom_rela}] may be rewritten as
\begin{equation}
a_{cons}(\Lambda,{\cal D}) - a_{cons}(\Lambda,{\cal D}')
+ \delta_\Lambda h({\cal D}',{\cal D})  =  0   , \label{gauge1}
\end{equation}
where $\Lambda$ is an infinitesimal generator of gauge transformation.
If we apply the second gauge transformation $\delta_{\Lambda'}$ to
Eq.[\ref{gauge1}],
and we subtract the result from the one which obtained by changing the order of
the transformations, we have
\begin{displaymath}
\delta_{\Lambda'}a_{cons}(\Lambda,{\cal D})
- \delta_{\Lambda}a_{cons}(\Lambda',{\cal D})
- \delta_{\Lambda'}a_{cons}(\Lambda,{\cal D}')
+ \delta_{\Lambda}a_{cons}(\Lambda',{\cal D}')
+ \delta_{i[\Lambda',\Lambda]} h({\cal D}',{\cal D})  =  0 ,
\end{displaymath}
where we have used $ [ \delta_{\Lambda'}, \delta_{\Lambda}] =
\delta_{i[\Lambda',\Lambda]}$.
Then, comparing this result with Eq.[\ref{gauge1}] again , one can easily show
that
\begin{equation}
\delta_{\Lambda'}a_{cons}(\Lambda,{\cal D})
- \delta_{\Lambda}a_{cons}(\Lambda',{\cal D})
= a_{cons}(i[\Lambda',\Lambda],{\cal D})  .
\end{equation}
This is nothing but the Wess-Zumino consistency condition.

Next let us consider the covariant anomaly. The covariant gauge anomaly
is given by \cite{witten}
\begin{equation}
a_{cov}(\Lambda,{\cal D}) = -  \, \lim_{\xi \rightarrow 0}
		\mbox{Tr}\, \gamma^{D+1} \Lambda \,
			e^{ - \xi {\cal D}{\cal D}} .
\end{equation}
Applying the relation $\delta_\Lambda {\cal D} = i [\Lambda, {\cal D}]$ to
Eq.[\ref{func}],
one can easily show that
\begin{equation}
a_{cov}(\Lambda,{\cal D}) = \delta_\Lambda
\widetilde{\Gamma}_{\rm o}[{\cal D}',{\cal D}]|_{_{{\cal D}' = {\cal D}}}
\end{equation}
where $\delta_\Lambda$ acts only on ${\cal D}$. Then from Eq.[\ref{two_var1}]
it is obvious that
\begin{equation}
a_{cov}(\Lambda,{\cal D}) = a_{cons}(\Lambda,{\cal D})
+ \delta_\Lambda h({\cal D}',{\cal D})|_{_{{\cal D}' = {\cal D}}} .
\label{cov1}
\end{equation}
Even if we add an arbitray invariant local function to $h$,
$a_{cov}$ of Eq.[\ref{cov1}] remains intact. The detailed proof of
this statement will not be given in this paper.
Since $ \delta_\Lambda A_\mu = [D_\mu, \Lambda] $,
 $\delta_\Lambda h({\cal D}',{\cal D})|_{_{{\cal D}' = {\cal D}}}$ can be
written in the form
 $ \int\,dx^D\, \det(e)\, \mbox{tr}\, [\Lambda,D_\mu]X^\mu $ with some local
function $X^\mu$.
So here we confirm again the arguement\cite{zumino} that the difference between
the consistent anomaly and the covariant anomaly comes from the diffrence
of the corresponding currents by some local function, i.e., by $X^\mu$.
Covariant gravitational anomalies can be also defined in a similar way.

The remaining part of this paper will be devoted to
the examples for the gauge anomaly. Since we are now interested only in the
gague anomaly,
we choose ${\cal D}_0
= -i E^\mu_{\ m}\gamma^m (\partial_\mu - \frac{i}{4} \omega^{mn}_\mu
\sigma_{mn})$.
Then  $D_\mu = \partial_\mu - i A_\mu - \frac{i}{4} \omega^{mn}_\mu \sigma_{mn}
$
	and $D'_\mu = \partial_\mu - i A'_\mu - \frac{i}{4} \omega^{mn}_\mu
\sigma_{mn} $.
When D=2, an unique solution of Eqs.[\ref{anom_rela}-\ref{anom_relb}] is
\begin{equation}
h(A',A) = C_1 \int dx^2\,\det(e)\,\epsilon^{\mu \nu}\, \mbox{tr}\, A'_\mu A_\nu
 , \label{dim2}
\end{equation}
where $C_1$ is a constant, which can be determined by comparing with the result
of
direct calculations.

When D=4 , we find
\begin{eqnarray}
h(A',A) & = & C_2 \int dx^4\,\det(e)\,\epsilon^{\mu \nu \lambda \tau} \times \\
\label{dim4}
	&   & \, \mbox{tr}\,
	\left( A'_\mu (A_\nu A_\lambda A_\tau + i A_\nu \partial_\lambda A_\tau
	   + i \partial_\nu A_\lambda A_\tau) - ( A \leftrightarrow A')
	   + \frac{1}{2} A'_\mu A_\nu A'_\lambda A_\tau \right),  \nonumber
\end{eqnarray}
where $C_2$ is a constant. From Eq.[\ref{anom_relb}] , we can see that $h$ does
not determine
$a_{cons}(\Lambda,A)|_{_{A=0}}$, in Abelian case, which is known to be
proportional to
$ i\int dx^4\,\det(e)\,\epsilon^{\mu \nu \lambda \tau} \,
\Lambda\,R_{\mu \nu \alpha \beta} R_{\lambda \tau}^{\,\,\,\,\,\alpha \beta}$.
With $h$ given above,  we can reconstruct the usual
  form of consistent chiral anomaly and covariant chiral anomaly.
We expect that more general cases of gauge and gravitation anomalies can be
investigated in a similar way.

Considering the functional determinants, we could derive an algebraic equation
which can be used to find the consistent and the covariant anomalies
simultaneously. Since our equation automatically satisfies
the Wess-Zumino consistency condition, all the considerations of
the consistency conditions are contained in Eq.[\ref{anom_rela}].
Eq.[\ref{anom_rela}]  is quite general, and therefore
can be applied to the cases of compact manifolds.

This work is supported in part by the Chungbuk National University,
and in part by the Ministry of Education in Korea.
The author wishes to thank Prof. M. Einhorn for helpful discussions
and for hospitallity at University of Michigan.

\end{document}